# Multiple Quintets *via* Singlet Fission in Ordered Films at Room Temperature

Daphné V. Lubert-Perquel[1], Enrico Salvadori[2,3*], Matthew Dyson[4,5], Paul N. Stavrinou[6], Riccardo Montis[1], Hiroki Nagashima[7], Yasuhiro Kobori[7], Sandrine Heutz[1*] and Christopher W. M. Kay[3,8*]

[1]*London Centre for Nanotechnology and Department of Materials, Imperial College London, Prince Consort Road, London SW7 2BP, U.K.*
[2]*School of Biological and Chemical Science, Queen Mary University of London, Mile End Road, London E1 4NS, UK*
[3]*Institute of Structural and Molecular Biology and London Centre for Nanotechnology, University College London, Gower Street, London WC1E 6BT, UK*
[4] *Department of Physics and Centre for Plastic Electronics, Imperial College London, London SW7 2AZ, UK*
[5]*Molecular Materials and Nanosystems and Institute for Complex Molecular Systems, Eindhoven University of Technology, P.O. Box 513, 5600 MB Eindhoven, The Netherlands*
[6]*Department of Engineering Science, Oxford University, Parks Road, Oxford OX1 3PJ, UK*
[7]*Molecular Photoscience Research Center, Kobe University, 1-1 Rokkodaicho Nada-ku Kobe, 657-8501, Japan*
[8]*Department of Chemistry, University of Saarland, 66123 Saarbrücken, Germany*

*The growing interest in harnessing singlet fission for photovoltaic applications stems from the possibility of generating two excitons from a single photon. Quantum efficiencies above unity have been reported, yet the correlation between singlet fission and intermolecular geometry is poorly understood. To address this, we investigated ordered solid solutions of pentacene in p-terphenyl grown by organic molecular beam deposition. Two classes of dimers are expected from the crystal structure – parallel and herringbone – with intrinsically distinctive electronic coupling. Using electron paramagnetic resonance spectroscopy, we provide compelling evidence for the formation of distinct quintet excitons at room temperature. These are assigned to specific pentacene pairs according to their angular dependence. This work highlights the importance of controlling the intermolecular geometry and the need to develop adequate theoretical models to account for the relationship between structure and electronic interactions in strongly-coupled, high-spin molecular systems.*

Interest in the photophysics of singlet fission (SF) has dramatically increased in recent years due to the possibility of overcoming the thermodynamic limitations in the efficiencies of organic electronic, organic spintronic and hybrid organic/inorganic structures[1–3]. SF is the photophysical process by which a single photon absorbed by a pair of chromophores generates two triplet excitons, with overall singlet character. These initially are strongly coupled triplets and may dissociate into free triplets or recombine via triplet-triplet annihilation. In its basic description, SF requires that the energy of the incoming photon is larger than twice the energy of the triplets generated. Being a spin-allowed process, it can be fast enough to outcompete prompt fluorescence. Harnessing this mechanism holds the promise to exceed the Shockley-Queisser limit[4] and quantum efficiencies >100% have been reported[5]. Only a few molecules are known to undergo SF, of which polyacenes are the most extensively investigated[6–11]. Recent studies identify the bound nature of the coupled triplets[10] and show evidence of the quintet nature of the coupled triplets in isotropic frozen solutions and amorphous films[11,12]. However, much remains unknown as to the precise nature of the multiexciton (paramagnetic) states and how the coupling between chromophores depends upon their geometry.

To address this, efforts have been devoted to the investigation of covalently bound chromophore dimers and van der Waals aggregates[13–15]. These provide excellent tunability of the electronic coupling but are impractical for real devices, as they are difficult to synthetize or are stable only under strictly controlled conditions. Spin-coating of pristine TIPS-tetracene and TIPS-pentacene provides a convenient medium to investigate the mechanisms of SF [11,12,16,17], but these disordered films require ultra-fast optics and/or cryogenic temperatures to study the population of spin states in order to suppress fast exciton diffusion[2]. Extensive theoretical and experimental work has been devoted to the description of singlet exciton properties of molecular aggregates, films and crystals, in terms of H- and J-type aggregates, but a recent study highlighted that charge transfer can introduce states beyond this traditional description. This is particularly the case for the herringbone structure adopted by polyacenes in the solid state[14]. However, analogous descriptions are lacking for the corresponding triplet excitons. As would be expected, local microstructure has been shown to play an important role in the dynamics of SF in pentacene compounds[7,9,18,19], as the physical arrangement of molecules in a solid determines the properties of excitons, including the extent of delocalization and associated relative admixture of charge-transfer configurations in the description of the exciton wave function[20]. Moreover, it has to be noted that the vast majority of SF studies are based on transient absorption spectroscopy due to its fs/ps time resolution at room temperature. Only recently has electron paramagnetic resonance (EPR) spectroscopy at cryogenic temperatures been used to selectively address triplet and quintet states that can be observed, but not distinguished, in optical/transient absorption experiments[11,12].

In order to allow detection and characterisation of the high-spin intermediates formed upon SF, we investigated ordered films of pentacene in a *p*-terphenyl matrix at different concentrations. Pentacene (Fig. 1a) was selected as the benchmark molecule for SF, whilst *p*-terphenyl (Fig. 1b) provides a well-defined host. The crystal structure of the pentacene in *p*-terphenyl has been documented – pentacene substitutes into one of two inequivalent sites, due to the herringbone structure of the host lattice[21–23].

Figure 1c depicts the energy levels and kinetics of the excited states in pentacene. Triplet states are populated *via* two distinct mechanisms that occur on different timescales and involve one or two pentacene units. SF is a charge-transfer (CT) mediated decay from $S_1$ to $^n$(TT), where *n* represents the spin species, which requires two interacting chromophores. These initial $^n$(TT) states, either dissociate into free triplets with uses in organic photovoltaics[1,2,5,24–26] or recombine into an emissive $S_1$ state. Alternatively, intersystem crossing (ISC) generates triplets *via* an $S_1$-$T_2$-$T_1$ transition. This is a single chromophore process, which is slower (nanosecond timescale) due to the spin-flip requirement, yet efficient in pentacene due to the quasi-resonant nature of the $S_1$ and $T_2$ states[27]. It follows that manipulating the concentration of chromophores allows control of the pathway in Scheme 1 even at ambient conditions. Therefore, EPR spectroscopy was used to characterise the progression from isolated to coupled photoinduced spin states in ordered films with in order to elucidate the effect of molecular aggregation and orientation on SF. Crucially, the conditions chosen for the experiments are those experienced by a real-life device, i.e. room temperature. To achieve these goals, films were prepared by organic molecular beam deposition. The rationale behind this was two-fold: (i) as co-evaporation is kinetically not thermodynamically driven, this technique enables pentacene to exceed its solubility limit and results in homogeneously doped films with no phase separation yet with defined structural ordering; (ii) precise control of pentacene concentration allows the progression from isolated molecules to van der Waals dimers and clusters to be studied. These points are significant as they result in localised long-lived excitons that can be studied at ambient conditions, a prerequisite to any functional material designed for devices.

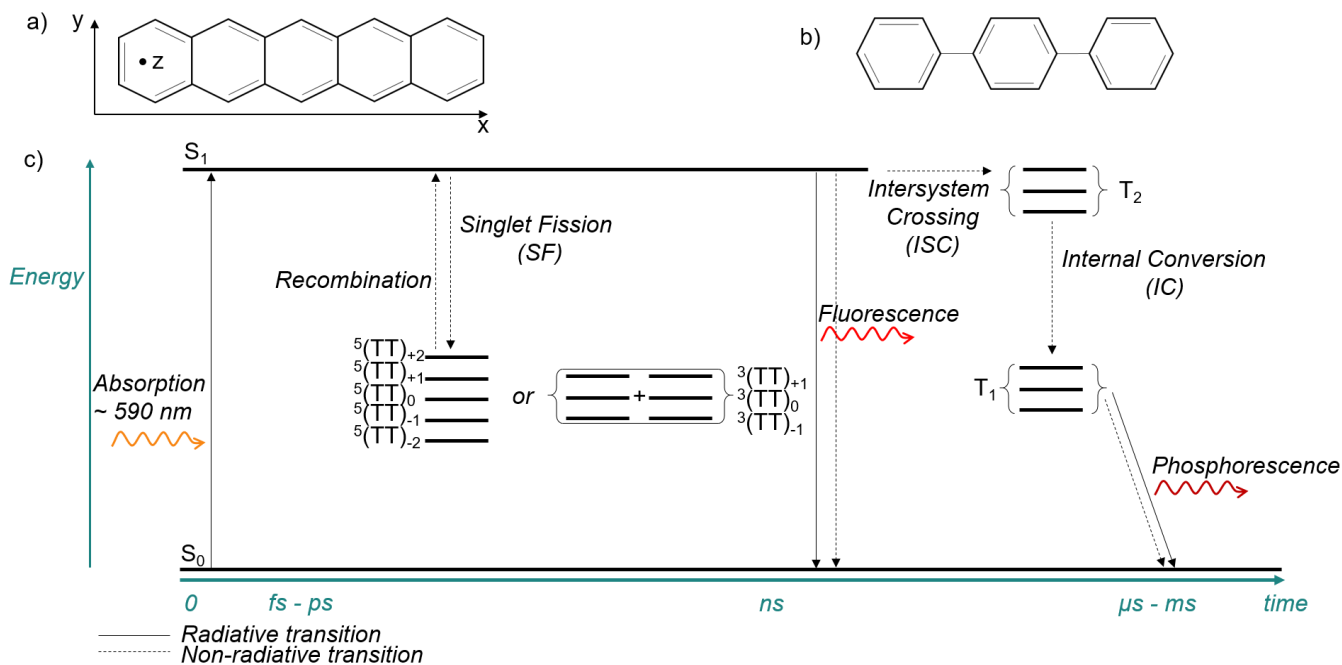

**Figure 1. Molecular structures and Jablonski diagram with timescale.**
*a, Molecular structure of pentacene with corresponding zfs axes. b, molecular structure of p-terphenyl. c, As a result of photoexcitation, pentacene generates triplets via two distinct mechanisms. Coupled pentacene molecules undergo very fast (fs-ps) SF leading to two strongly coupled (TT) states of overall triplet or quintet nature. These can recombine or ultimately dissociate into isolated triplets. Alternatively, individual pentacene molecules can transition to triplet states via ISC, which is energetically less favourable due the required spin-flip and therefore occurs on slower timescales (ns).*

**Morphology and structure**

Pentacene and *p*-terphenyl are aromatic molecules which adopt the triclinic and monoclinic crystal systems respectively. Films of 1μm thickness were deposited by co-evaporation. The pentacene molecule substitutes into the *p*-terphenyl lattice, replacing a host molecule in one of two sites within the herringbone structure of the host[23].

The film morphology is shown in Figure 2a and Figure S2. In very dilute samples, the films form micron sized platelets comparable to those of pure *p*-terphenyl, shown in Figure S1. From 1% dopant concentration, small grains approximately 50 nm in diameter appear within the crystallites. Above 10%, the platelets fragment into smaller grains which is exacerbated with increasing concentration and at 50% the crystallites are altogether more faceted.

X-ray diffraction (XRD) indicates that the structure of dilute samples (<1%) is unchanged and a single sharp diffraction peak corresponding to texture along the (003) plane of *p*-terphenyl is observed (Fig. 2c). However above 1%, broadening is observed as well as a linear peak shift towards lower scattering angles. The peak broadening indicates decreased crystalline coherence length, predominantly attributed to reduced grain size (apparent from SEM) via the Scherrer equation[28] (Fig. 2d), although lattice strain induced disorder may also contribute. Due to the substitution of larger molecules within the host, a gradual increase in the lattice parameter (i.e. reduced scattering angle) with increased pentacene concentration is observed, a trend that also occurs in solid solutions of inorganic materials[29–31]. Furthermore, there is no evidence for phase separation either in XRD or SEM, implying a homogeneous distribution of molecules within the films.

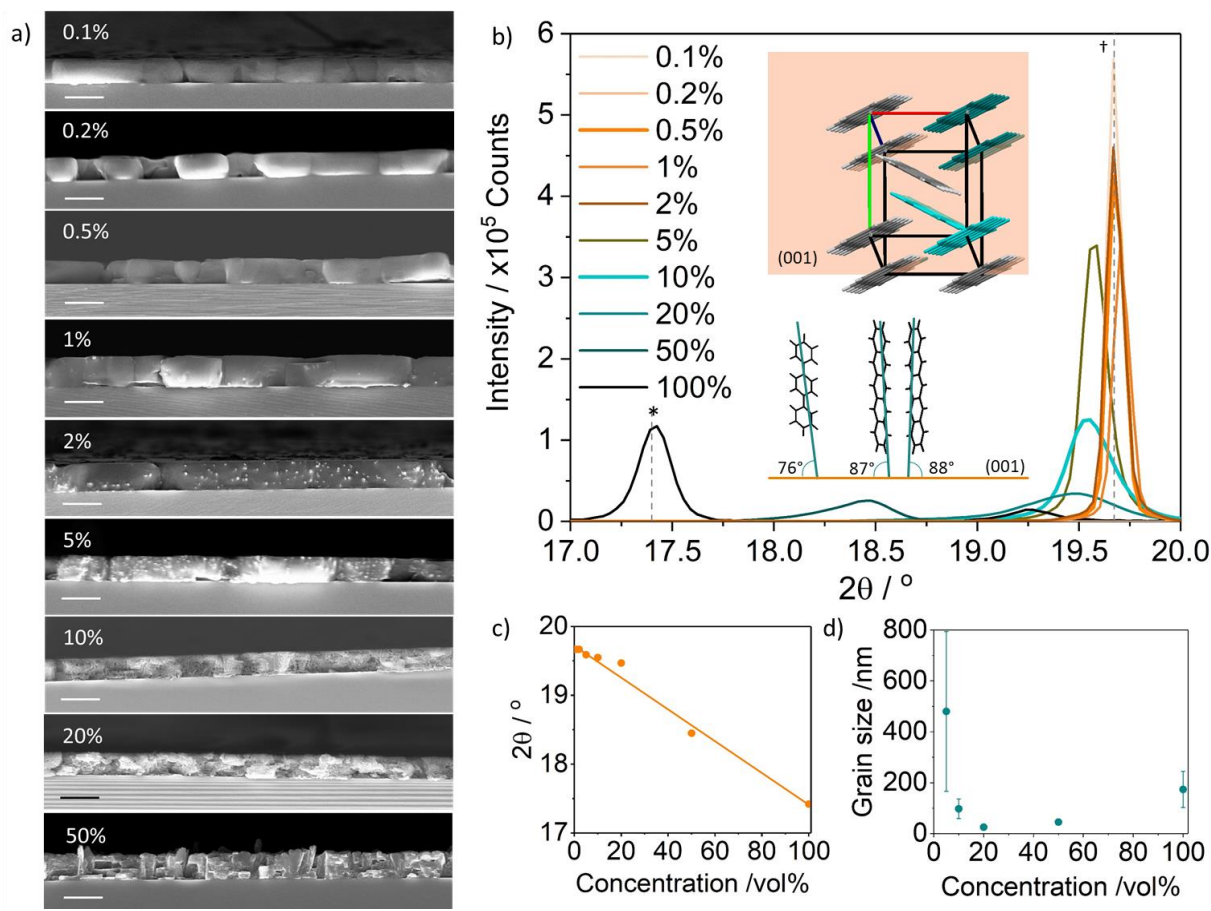

*Figure 2. Morphology and structure of the 1µm thick pentacene doped p-terphenyl films as a function of concentration. **a**, SEM cross-sections with a 1µm scale bar. The very dilute samples form large platelets typical of p-terphenyl whereas an increase in dopant concentration leads to fragmentation into smaller faceted grains. **b**, Close-up of the (003) peak where (\*) corresponds to the pentacene peak position and (†) p-terphenyl position. The top inset shows the unit cell of the pentacene with the parallel dimer in teal and herringbone dimer in cyan. The second inset shows the contact angle of p-terphenyl (left) and pentacene (right) on the substrate. **c**, The peak position shifts linearly towards lower scattering angles indicating increased lattice spacing. **d**, Shows the variation in grain size with pentacene concentration calculated from the Scherrer equation. 5% and 100% pentacene are very close to instrument resolution and consequently have the largest error. All films are 1µm except the 100% pentacene which is 100nm – the XRD intensity is normalised to the thickness.*

**Optical determination of aggregation**

Absorption spectra were measured as a function of concentration to characterise aggregate formation (Fig. 3a). All optical measurements were performed in an integrating sphere to account for the effects of scattering, which is not consistent across all concentrations due to the topological and structural variations detailed above. Below 1% pentacene in *p*-terphenyl, the absorption spectra are distinctly those of single molecule pentacene. The lowest energy peak, at 590 nm, corresponds to the $S_0 – S_1$ transition with the associated vibronic progression at higher energies. Above this threshold a longer wavelength shoulder appears due to Davydov splitting of the lowest energy transition[32], suggesting pentacene aggregate formation – the concentration of these species would be below the XRD sensitivity. In more concentrated samples, there is an additional contribution at higher energies (~ 540, 580 nm) due to absorption from $S_0$ to two pentacene-pentacene charge transfer (CT) states[33]. These are generated in π-stacks with strong intermolecular interactions and nearest neighbour separations on the order of ~3.5 Å [14,34,35]. Frenkel and CT exciton mixing, mediated by delocalised excitons, have been suggested to lead to J- or H- charge transfer aggregates unlike the classical Kasha model[14,36].

Photoluminescence quantum efficiency (PLQE), indicative of the proportion of absorbed photons that subsequently decay radiatively, shows an anticorrelation with absorption at the excitation wavelength of 590 nm, up to 5% pentacene concentration – above this proportion, fluorescence is quenched below detector sensitivity (Fig 3b). As we demonstrate below, this is attributed to the faster singlet fission process supplanting fluorescence.

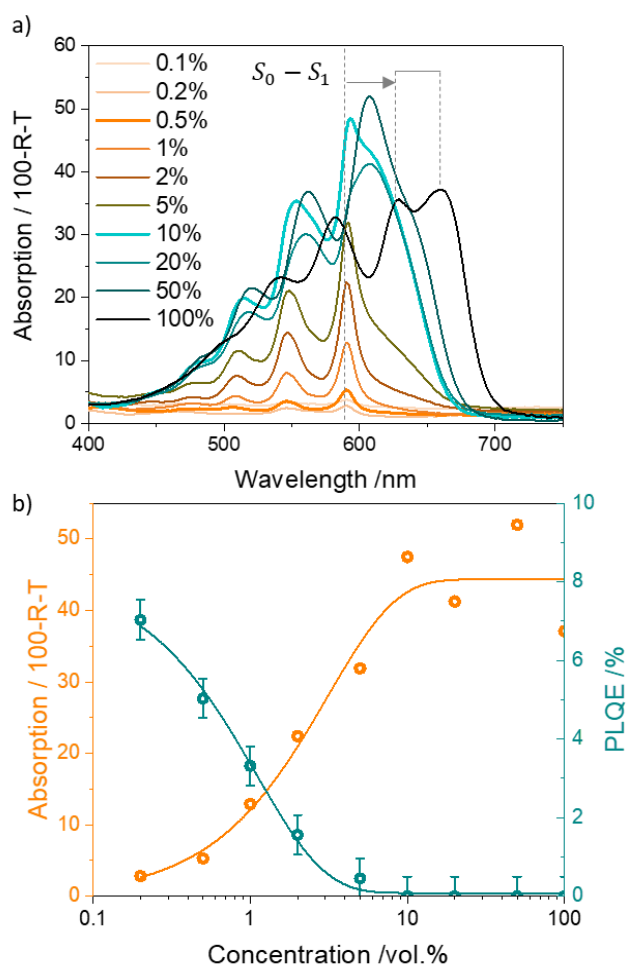

*Figure 3. Optical spectroscopy of the pentacene doped micron films.*
*a, Absorption spectra of the films as a function of pentacene concentration, with reflection and transmission subtracted from incident intensity (100-R-T). Below 2% pentacene the single molecule spectra is seen with the $S_0 – S_1$ transition at 590 nm, while the Davydov split peak gradually appears above this concentration. A noticeable red-shift is also observed at higher concentrations. b, The anti-correlation of absorption at 590 nm and photoluminescence quantum efficiency (PLQE) measured at 590 nm excitation, as a function of concentration – trend lines are a guide to the eye. Above 10% pentacene peak shifts affect the extracted value, whilst PLQE is too low to detect.*

**Orientation of pentacene molecules in the film**
XRD characterisation did not enable firm conclusions to be drawn about the orientation of pentacene in the host matrix. However, this could be achieved by exploiting the anisotropy of zero-field splitting interaction using time-resolved EPR spectroscopy (tr-EPR)[37]. The full rotation pattern and simulation are shown in the supplementary information (Fig. S4). The analysis allows us to conclude that the pentacene lies with the x-axis at 70° ± 5° with respect to the substrate, i.e. matching the orientation of the *p*-terphenyl molecules with no in-plane preferential orientation.

The paramagnetic excited states of the pentacene molecules were also investigated as a function of concentration and the results for the molecular x-axis (Fig. 1a) parallel to the magnetic field are shown in Figure 4a. The preferential orientation is maintained up to 20% pentacene, above which the EPR signal is quenched, suggesting clusters of pentacene are sufficiently close to enable exciton hopping at room temperature. As above, the outer peaks can be accounted for by an ISC triplet contribution whereas the inner peaks in the shaded region, which are consistently observed above 1%, do not correspond to a texture variation but to a different spin species.

## Multiexciton spin-states

Figure 4b depicts a 2D mapping of the TREPR spectrum of the 10% pentacene: p-terphenyl blended film with the applied magnetic field parallel to the x-axis. Four pairs of peaks are visible, with the two inner peak separations about 1/3 of the outer peak separations. This suggests two pairs of strongly exchanged coupled triplets (J > D), forming quintet states with S = 2 [11,12]. The quintet lifetime of the innermost peaks (at 334 and 349 mT), which is most apparent in the emissive line, is ~300 ns, and its decay coincides with a rise in the corresponding triplet signal (at 315 and 369 mT). This implies a short-lived high-spin state that efficiently separates into free triplets, which in turn are very long-lived (> 3 μs). On the other hand, the second pair of quintets (at 331 and 351 mT) has lifetime of ~1 μs and appears to coexist with the weak, outermost triplet signal (at 312 and 372 mT). This could result from a thermally activated equilibrium between the quintet and triplet states. The experimental time traces are shown in the supplementary information, Figure S8.

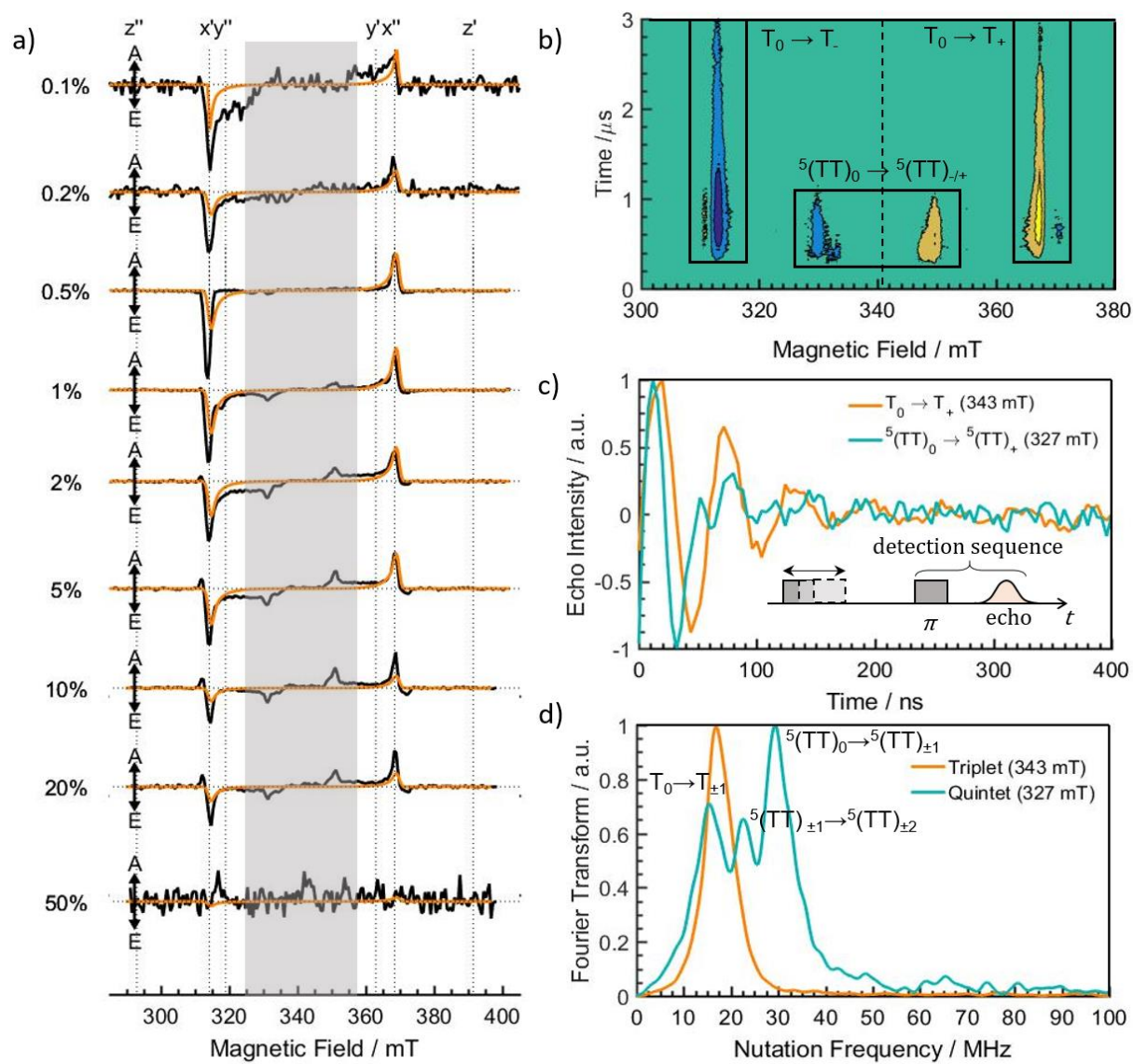

*Figure 4. Identifying quintet states at room temperature via EPR spectroscopy and nutation experiments of the (TT) and isolated T states.*
*a, EPR spectra acquired in the x-orientation as a function of pentacene concentration with the simulated data shown in orange. The inner peaks are assigned to long-lived strongly coupled triplet states. b, Tr-EPR of the 10% pentacene in p-terphenyl 1μm film. The $^5$(TT) states (inner peaks), appear earlier in time than the isolated T states (outer peaks), formed by dissociation or ISC. c, Nutation experiment for the high-field quintet and triplet transitions, with the pulse sequence shown in the inset. d, Fourier Transform of the data showing a ratio of 1.74 between the triplet and quintet nutation frequencies, in accord with the nominal value of √3 = 1.73.*

The most conclusive way to determine the spin state of the species giving rise to the peaks observed in the tr-EPR surface is to perform transient nutation measurements at the relevant field positions[11,12]. The observed frequencies directly depend on the total spin (S) and on the $m_S$ component and therefore allow for a definitive assignment of the different transitions. The explicit formula correlating the nutation frequency ($\omega_{m_S,m_{S\pm1}}$) to the total spin (S) and the spin projections characteristic of the transition observed ($m_S$ and $m_{S\pm1}$) reads:

$$\omega_{m_S,m_{S\pm1}} = \omega_{1/2}\sqrt{S(S+1) - m_S(m_S \pm 1)}$$

where $\omega_{1/2} = g\mu_B B_1/\hbar$ is the precession frequency for a spin S = 1/2 and $B_1$ is the microwave field strength. It follows that for transitions from the $m_s$ = 0 state, the ratio of nutation frequencies for pure quintet and triplet states is $\sqrt{3}$. Conversely, for transitions from the $m_S$ = 1 the ratio equates to $\sqrt{2}$. Comparison between measurements at different field positions but with all the other experimental settings unchanged (note that nutation frequencies depend on the strength of $B_1$ and therefore on the Q factor) provides a way of probing the spin state of the observed transitions. Figure 4c,d shows the nutation frequency and corresponding Fourier transform of the long-lived triplet and quintets at the high-field position. The FFT shows the triplet and corresponding quintet nutation frequency with a ratio of $\sqrt{3}$, as well as an intermediate peak with ratio $\sqrt{2}$. The former contribution is an $S_0$ to $S_{\pm1}$ transition and the latter is a $S_{\pm1}$ to $S_{\pm2}$ transition. This indicates the high spin states of the quintet are also populated in this system.

In order to disentangle the individual contributions and to estimate the magnetic parameters of the excited spin states the x- and yz-orientations were simulated simultaneously with a single set of parameters, detailed in the supplementary information. The experimental and simulated spectra for the 10% pentacene film are shown in Figure 5, for both the x-orientation and the y,z-orientation (20° and 110° ±5° with respect to the B-field). As predicted from the host crystal structure, the nearest neighbours are separated by ~4 Å edge-to-edge and can either be in a herringbone or parallel geometry. The proximity of adjacent pentacene molecules results in strong coupling, which in principle makes singlet fission possible for both conformations. Therefore, the simulations included three contributions: 1) a pair of interacting triplets representative of the parallel conformation (teal dimer in the inset of Fig. 2b); 2) a pair of interacting triplets representative of the herringbone conformation (cyan dimer in the inset of Fig. 2b); and 3) ISC triplet stemming from isolated pentacene units still present at 10% doping. In the x-orientation, the simulation used an isotropic J = 20 GHz, with an added spin-spin interaction component of 60 MHz for the parallel dimer configuration, as calculated in previous studies[11,38]. It should be noted the isotropic J-coupling stated is a minimum threshold not a precise value.

The angle between pentacene molecules in the herringbone dimer configuration was set to the average of the host and dopant lattice parameters as an exact crystallographic structure has not yet been determined. The most precise description of this solid solution to date is obtained through tr-EPR and is presented in the supplementary information. The added angular contribution to the dipolar coupling accounted for the phase change between the quintet and triplet peaks of the EPR spectra whereas the collinear coupling resulted in the same triplet and quintet phase. The outer pair of triplets and corresponding quintet could therefore be attributed to the herringbone dimer and the inner pairs to the parallel dimer. Dimerization has been shown to alter the zero-field splitting (zfs) parameters with respect to the constituent monomers although no clear link to morphology has been made[13,39]. Here, we conclude that the parallel dimer has a zfs unchanged with respect to the monomer whereas the zfs increases in a herringbone conformation.

In the yz-orientation, the in-plane disorder added complexity to the model. The simulation reported in Figure 5 used a 90° rotation of the x-orientation calculation, keeping all coupling parameters the same and the peak positions were correctly attributed to y,z molecular orientations. However, the two innermost peaks correspond to an x-orientation quintet, following the trend that increased concentration leads to increased disorder. An S = 2 model for a polycrystalline sample was also calculated (Fig. S7), which includes an ordering parameter that could account for the increased texture. With this method, the full spectrum was simulated but no precise J-value was computed. Table 1 reports a comparison with literature data on analogue systems. The magnetic parameters derived here for ordered films compare satisfactorily with those obtained from amorphous films and frozen solutions.

*Table 1. Comparison of the fitted parameters derived in this work with previously reported data on analogue systems.*

| Sample | Form | Temperature (K) | $J_{iso}$ (GHz) | X (MHz) | g | |D| (MHz) | |E| (MHz) | Distance (Å) | Ref. |
|---|---|---|---|---|---|---|---|---|---|
| TIPS-tetracene | Amorphous film | < 75 | $1.02 \times 10^3$ $^\$$ | | | | | | [12] |
| Pentacene dimer (PB2) | Frozen Solution | 80 | 29.2 | 10 | 2.001 | 1078 | 13 | | [11] |
| Pentacene dimer (PB3) | Frozen Solution | 20 < T < 80 | 19.9 | 39 | 2.002 | 1138 | 19 | 16 | [11] |
| Pentacene | Oriented film | 298 | >20 | - (*) 60 | 2.0023 | 1400 1600 | 50 80 | ~ 3-4 | This work |

*(\*) herringbone dimer is treated as isotropic and parallel dimer has large anisotropy.*
$^\$$ *Estimated upper limit*

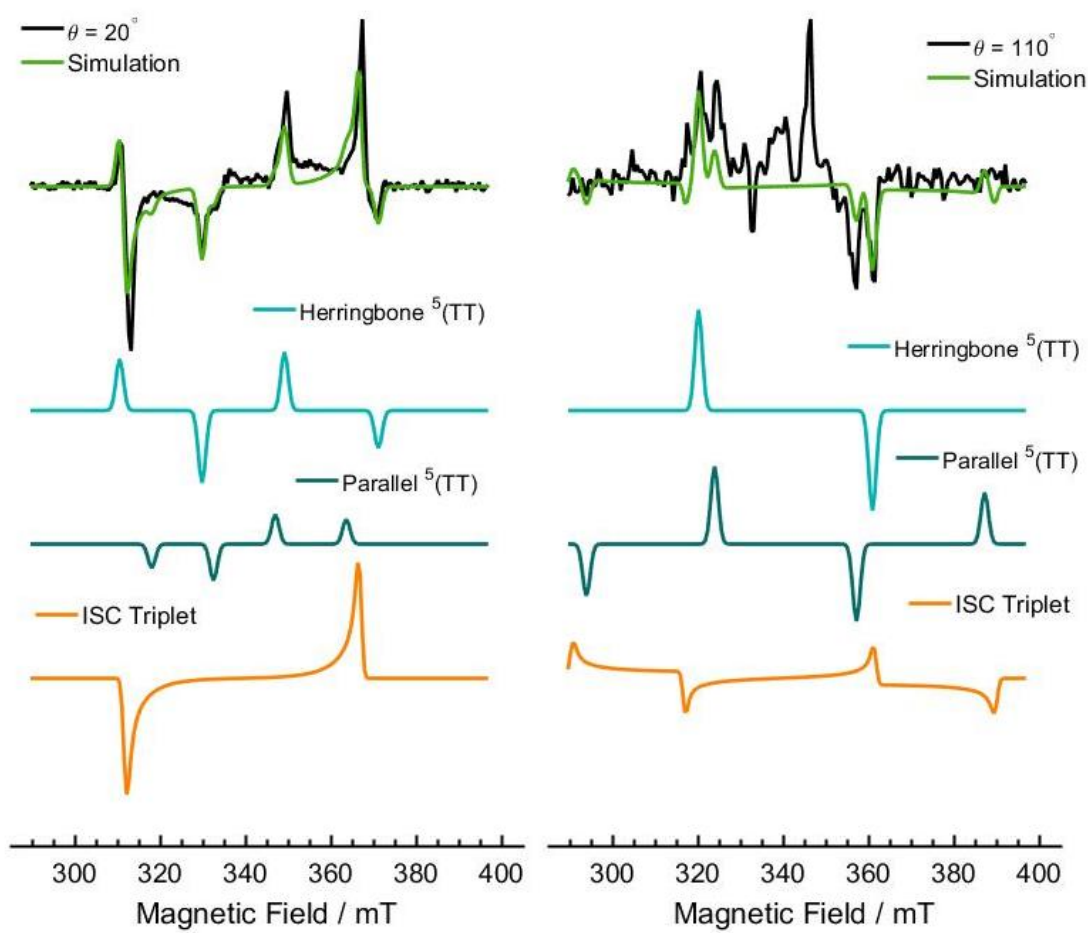

*Figure 5. Evidence of multiple spin species resulting from SF for the 10% pentacene doped p-terphenyl micron film at RT. The experimental spectra (black) can be fitted (green) with a superposition of simulated peaks for the herringbone (cyan) and parallel (teal) dimer arrangement of strongly coupled triplets as well as the intersystem crossing triplet (orange).*

## Conclusions

Singlet fission leads to the generation of excited triplet and quintet states on pairs of interacting chromophores. Only recently quintet formation has been observed by EPR spectroscopy at cryogenic temperatures from pentacene dimers in solution and TIPS-tetracene cast as amorphous films. Indeed, EPR spectroscopy is the sole technique able to unambiguously identify and distinguish between triplets and quintets. Two features provide the signature of a quintet states: the peak splitting in the EPR spectrum three times smaller than that of the corresponding triplets and the nutation frequency $\sqrt{3}$ times larger than that of free triplets. A major ambiguity in the understanding of singlet fission stems from the poor experimental correlation between molecular geometry and spin coupling. We exploited our growth method to bridge the gap between covalent dimers in solution and monomers in films but crucially at room temperature and with highly controlled molecular orientation. Remarkably, the use of ordered structure and a progressive increase in pentacene concentration allowed the unprecedented observation and characterisation of two strongly coupled quintet states. Moreover, dilution inhibits excitation/spin diffusion, significantly enhancing the excitation lifetime. We have provided direct evidence for two distinct spin coupled states that can be assigned to pairs of pentacenes on the basis of their relative orientations. This unambiguously proves that the structure of the pentacene dimers directly affects the properties and dynamics of the corresponding triplet and quintet high-spin states and that specific geometries are likely to promote the efficiency of SF and to extend the lifetime of the triplet excitons.

## Acknowledgements

DLP acknowledges a studentship from the EPSRC CDT for the Advanced Characterisation of Materials (EP/L015277/1) and a research placement supported by the joint research program of Molecular Photoscience Research Center, Kobe University, under the supervision of YK. MD thanks the UK Engineering and Physical Sciences Research Council (EPSRC) Plastic Electronics Doctoral Training Centre (EP/G037515/1) for funding.

## Author Contributions

RM purified the pentacene and p-terphenyl. The films were fabricated and characterised by DLP. Optical spectroscopy was carried out by MD and DLP under the supervision of PNS. MD performed photoluminescence calculations. DLP and ES carried out EPR spectroscopy measurements. DLP and ES interpreted the experimental results. YK and HN provided insights into the formalism of the SF mechanism. DLP prepared the figures. DLP and ES wrote the manuscript with critical feedback from all authors. The study was conceived and supervised by SH and CWMK.

**Methods**

**Thin film fabrication**. Commercially obtained pentacene purified by sublimation (TCI UK Ltd) and *p*-terphenyl (99+% Alfa Aesar) further zoned refined were deposited by organic molecular beam deposition using a Kurt J. Lesker SPECTROS 100 system with a base pressure $10^{-7}$ mbar. Quartz crystal microbalance sensors were used to precisely control the deposition rates and therefore the concentration of the thin films. The concentrations are expressed as the ratio by volume of pentacene relative to *p*-terphenyl with the precise values to within 0.1% as follows: 0.1%, 0.2%, 0.5%, 1.1%, 2.0%, 4.8%, 9.0%, 16.8%, 50.2%, 100%. For simplicity, these are referred in the main text as: 0.1%, 0.2%, 0.5%, 1%, 2%, 5%, 10%, 20%, 50% and 100%.

**Scanning Electron Microscopy.** The morphology was investigated using a high-resolution LEO Gemini 1525 field emission gun scanning electron microscope (FEGSEM). To obtain the cross-sections, thin films deposited on Si(100) were cleaved to obtain a sharp surface.

**X-ray Diffraction.** For the structural measurements, a PANalytical X'Pert Pro multi-purpose diffractometer was used with a Cu K$_\alpha$ X-ray source. The grain size was calculated using the Scherrer equation, with the difference in broadening between a Gaussian and Lorentzian fit used as the uncertainty in calculation[40]. Large grains have a much larger error as these are at the instrument limit.

**Absorption and PLQE.** Absorption spectra were determined using a Shimadzu UV-2600 spectrometer fitted with an integrating sphere, enabling scattered light to be collected, thus removing this loss contribution from the absorption spectra. Furthermore, the total reflection (i.e. diffuse and specular) was measured, using a negligibly absorbing fused silica substrate as a reference, and subtracted along with transmission from 1 (i.e. total incident intensity), leaving only absorption.

Photoluminesence spectra and the corresponding PLQE values were determined by placing the thin film sample within a 15 cm diameter integrating sphere (Labsphere, internally coated with Spectraflect), before excitation using a monochromated supercontinuum Fianium light source at 590 nm. Both excitation and emission were transferred using a 100 μm diameter optical fibre to an Andor SR-163 spectrometer and recorded with an Andor i-Dus CCD. Integrating sphere reflectivity and CCD response was corrected for using a calibrated halogen light source (HL-2000-CAL, Ocean Optics). PLQE spectra was determined using the de Mello method[41], by comparing the intergrals of excitation (attenuated by absorption) and emission peaks for direct and indirect excitation along with the excitation peak with an empty sphere. PLQE error bars are estimated at 0.5%.

**EPR spectroscopy**
All EPR measurements were acquired using a Bruker E580 pulsed EPR spectrometer operating at X-band frequencies (9–10 GHz/0.3 T), operating either in continuous wave or pulsed mode, equipped with an Oxford Instruments CF935O flow cryostat used to support a Bruker ER4118-X MD5 resonator.
A Surelite broadband OPO system within the operating range 410–680 nm, pumped by a Surelite I-20 Q-switched Nd:YAG laser with 2nd and 3rd harmonic generators (20 Hz, pulse length: 5 ns) was used to achieve a pulsed laser excitation at 580 nm, the maximum absorption for pentacene, with the energy at the sample approximately 5 mJ per pulse.
All spectra were recorded at room temperature in air atmosphere.

**Time-resolved EPR Spectroscopy**
Time- resolved EPR (tr-EPR) spectra were recorded in direct detection mode without magnetic field modulation. Hence, they show characteristic enhanced absorptive (A) and emissive (E) features. The rotation patter, used to derive the orientation of pentacene molecules within the sample, was recorded by means of a laboratory-built goniometer in step of 10 ± 2°. The sample was mounted on a bespoke quartz support designed to keep the film slide upright with the long axis parallel to the axis of the resonator. TR-EPR spectra were simulated using the function *pepper* present in the EasySpin[42] toolbox running on MATLAB™. Oriented spectra were calculated as a superposition of spectra of the two inequivalent sites, related by crystallographic (P $2_1$/a), taking as starting values those reported for bulk pentacene:*p*-terphenyl crystal at room-temperature[43,44].
The simulation parameters include the zfs tensor principal values and the orientation of zfs for one site in the unit cell, the orientation of the crystal in the magnetic field, and a line width parameter that accounts for the unresolved hyperfine interactions. Given that the zfs tensor is traceless, its principal values $D_X$, $D_Y$ and $D_Z$ can also be expressed by two zero-field splitting parameters $D$ and $E$. These are related to the principal values according to the relations: $D_X=1/3D+E$, $D_Y=1/3D-E$ and $D_Z=-2/3D$. zfs parameters and sublevel populations were taken from literature values, with D = 1400 MHz, E = 50 MHz [45] and $P_x$:$P_y$:$P_z$=0.76:0.16:0.08 [46]. An isotropic *g*-value equal to the free electron *g* value ($g_x = g_y = g_z$ = 2.0023) was used in all simulations.

**Pulsed EPR spectroscopy**
Echo-detected field sweep spectra (not shown) were recorded using the pulse sequence π/2-200-π-200-echo with a π/2 pulse of 16 ns. Transient nutation spectra were recorded at fixed magnetic fields with the shortest delay after flash available ~ 100 ns and pulse sequence $P_{nutation}$-400-π-400-echo, with $P_{nutation}$ starting at 4ns and incrementing by 4ns for 101 points. The resulting echo was integrated in quadrature detection as a function of nutation pulse length, $P_{nutation}$. The time-domain data were baseline corrected with a second order polynomial function, tempered with and Hamming window function, zero-filled and Fourier transformed to establish the nutation frequency of a given transition. The absolute part is reported in the figure.